\documentclass[a4paper,11pt]{article}
\usepackage{graphicx}
\pdfoutput=1 

\usepackage{jheppub} 

\usepackage[T1]{fontenc} 

\title{\boldmath Phase transitions and Geometrothermodynamics of Regular black holes}

\author[1]{R. Tharanath,\note{Corresponding author.}}
\author{Jishnu Suresh,}
\author{V. C. Kuriakose}

\affiliation{Department of Physics,Cochin University of Science and technology\\Cochin, Kerala, India}

\emailAdd{tharanath.r@gmail.com}
\emailAdd{jishnusuresh@cusat.ac.in}
\emailAdd{vck@cusat.ac.in}

\abstract{In this paper we study the thermodynamics and state space geometry of regular black hole solutions such as 
Bardeen black hole\cite{Bardeen}, Ay\'{o}n-Beato and Garc\'{i}a 
black hole\cite{abg1}, Hayward black hole\cite{Hayward} and
Berej-Matyjasek-Trynieki-Wornowicz black hole\cite{bereg2}. We find that all these black holes show second order thermodynamic phase transitions(SOTPT) by observing  
discontinuities in heat capacity-entropy  graphs 
as well as the cusp type double point in free energy-temperature graph. Using the formulation of geometrothermodynamics we again find the singularities in the heat capacity of the black holes
by calculating the curvature scalar of the Legendre invariant metric.}

\begin{document} 
\maketitle
\flushbottom

\section{Introduction}

The realization of the fact that, the four laws of black hole mechanics are identical to the four laws of thermodynamics span the platform of black hole 
thermodynamics. In thermodynamic studies we are mainly interested in
the response of certain thermal quantities of the system towards change of temperature. If there exists any abnormal behavior in these quantities, it requires a
detailed study. If we observe a discontinuity in heat capacity as the temperature of the system is varied, it is considered that the system has undergone 
a phase transition. 
It is found that the ordinary black holes show second order thermal phase transitions.  Most of these black holes are thermodynamically unstable 
for certain ranges of temperature.

One of the major problems of general theory of relativity is that the theory possesses singular solutions.  A black hole is usually identified
as a space time having singularity at the origin enclosed by event horizon. Hence, black hole solutions are the best examples showing singular
behavior in general theory of relativity.
But in 1968, Bardeen\cite{Bardeen} obtained a black hole solution without any singularity at the origin possessing an event horizon. Later many have obtained
similar solutions \cite{borde1,borde2,abg1}. The coupling of non-linear electromagnetic theory to 
general relativity has brought to new sets of charged black holes. Certain black holes came in the range of regular black hole solutions as well.
 In 1999, Ayon Beta and Garzia\cite{abg1} found such a black hole solution. Later in 2006, Berej et. al.\cite{bereg2} as well as Hayward\cite{Hayward} found different
kinds of such black hole solutions. 
These black holes are in general called 'regular' or 'non singular' black holes. 

Various aspects of regular 
black holes like geodesic
motions of test particles around regular black holes \cite{Z}, gravitational lensing \cite{E}, gravitational and electromagnetic stability \cite{mor,fer},
thermodynamics of regular black holes\cite{akbar} were studied earlier.  We here study the thermodynamic behavior of regular black hole.
Looking at the thermodynamic stability of these black holes, we observe that all of these black holes exhibit second order thermodynamic phase transitions.
But in the case of ordinary black holes this is not always observed.
In the case of Schwarzschild black hole, the heat capacity is found to be negative\cite{tharanath,tharanath2}. 
It readily implies that the thermal emissions from the black holes will definitely grow in an uncontrolled manner and that black hole will 
end up in an unstable thermodynamic state. In the case of Reissener-Nordstrom black hole, it has got both positive and negative heat capacities. 
In this case the extra parameter              , the charge, drives this black hole from an unstable state to stable state via a second order thermodynamic phase transition.

Hermann\cite{hermann} and Mrugala\cite{mrugala} introduced a geometric approach to study thermodynamics. Weinhold\cite{weinhold} 
introduced another method with a metric which has been widely used 
to study the physical properties of various thermodynamic systems and the associated Riemannian structure. 
Later Ruppeiner\cite{ruppeiner} introduced a metric which is conformally equivalent
to Weinhold's metric. But these two methods entirely depend on the choice of thermodynamic potentials. 

Geometrothermodynamics(GTD)\cite{q1,q2,q3} is the latest attempt in this direction. In this approach the metric is built up from a Legendre
invariant thermodynamic potential and from its first and second order partial derivatives with respect to the extensive
variables. The earlier studies show that the thermodynamic stability of systems depend on the potential we have chosen.
This contradiction can be removed by using new Legendre invariant metric as introduced in the GTD. Above all, 
GTD has a unique feature in identifying the second order phase transition point\cite{tj1,tj2}, among the geometric methods we have described
above. In this method the phase transition can be described in terms of curvature singularities.
Phase transitions shown by black holes are the main
thermodynamical attraction in the present study. And such phase transitions can be well studied with the behaviour of free energy of the black hole.
Here we are looking for a correct geometric explanation of phase transitions given by these black holes.

The present work is organized as follows. In section 2,  we review the thermodynamics of regular black holes such as Bardeen black hole,Ay\'{o}n-Beato and Garc\'{i}a 
black hole, Hayward black hole and Berej-Matyjasek-Trynieki-Wornowicz black hole. In section 3, we analyse the Geometrothermodynamics of these regular black holes.
Final section is devoted for conclusion. 
\begin{figure}[h]
\centering
\includegraphics[scale=0.65]{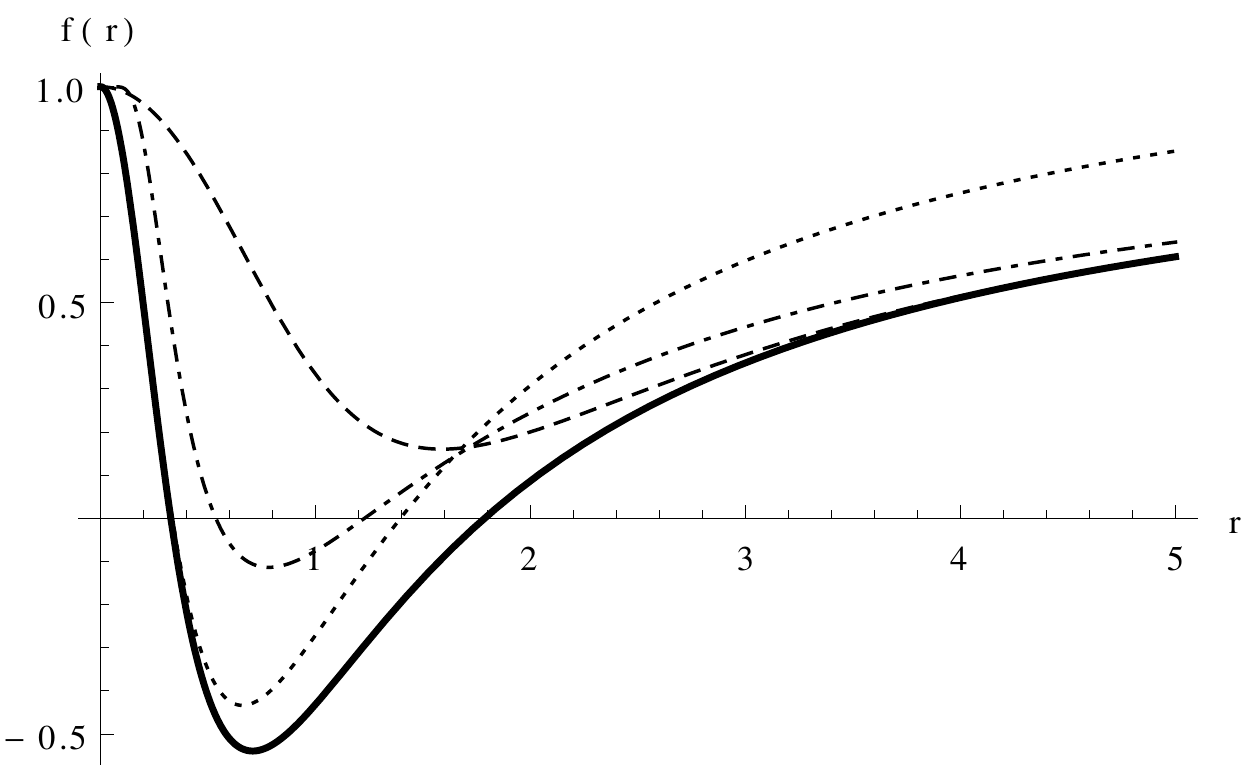}
\caption{Variation of $f(r)$ functions of the four regular black holes, Bardeen(thick), ABG(dotted), Hayward(dashed) and BMTW(dot-dashed) together have 
plotted here, for $r_+$.}
\label{forall}
\end{figure}
\section{Thermodynamics of regular black holes}

\subsection*{Bardeen black hole.}

The possibility of the existence of regular solutions of Einstein equation is very much restricted because of various singularity theorems which apply both 
to the cosmology and gravitational collapses. It becomes interesting to see that event horizons could be formed without a singularity at the origin. 
Singularity of black hole is an acknowledged difficulty in general relativity. At the singular point, the curvature will be divergent, 
so it means all the physics laws fail at the point. 
The first successful attempt in this direction was made by Bardeen \cite{Bardeen} in 1968 who obtained a solution of Einstein's equation with an event 
horizon but without a singularity at the origin. Bardeen's solution of Einstein's equation in the presence of nonlinear electromagnetic field is parametrized 
by mass $M$ and charge $q$. The static and spherically symmetric line element is given by
\begin{equation}
ds^{2}=f(r)dt^{2}-\frac{dr^{2}}{f(r)}-r^{2}(d\theta^{2}+\sin^{2}\theta d\varphi^{2}),
\end{equation}
where

\begin{equation}
f(r)=1-\frac{2Mr^{2}}{(r^{2}+q^{2})^{\frac{3}{2}}},
\end{equation}

and this function is well defined everywhere for $r\geq 0$. The value of $\frac{M}{q}$ can be chosen that $f(r)$ is always positive 
and it has two zeros. 
The graph (Fig\ref{forall}), $f(r)$ vs $r$ reveals this fact. Using the area law($S=4\pi r^2$) we can write the mass in terms of the entropy $S$ and the $q$ as

\begin{figure}[h]
\centering
\includegraphics[scale=0.65]{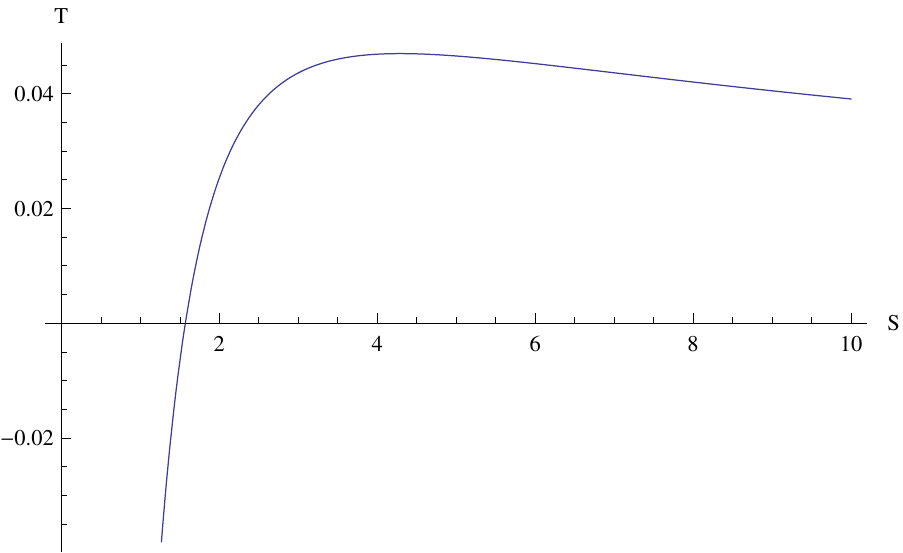}
\caption{Variation of temperature with respect to entropy $S$ of Bardeen black hole.}
\label{tempbardeen}
\end{figure}

\begin{equation}
M= \frac{( S+\pi q^2)^{\frac{3}{2}}}{2 \sqrt{\pi}S}.
\end{equation}

\begin{figure}[h]
\centering
\includegraphics[scale=0.65]{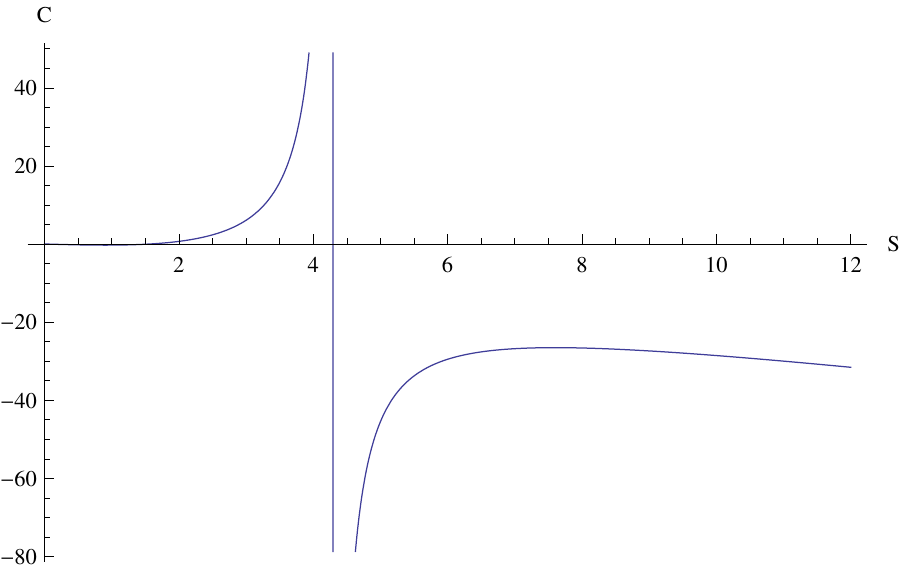}
\caption{Variation of heat capacity with respect to entropy $S$ of Bardeen black hole.}
\label{heatbardeen}
\end{figure}
Temperature of the black hole is given by the relation, $T=\frac{\partial M}{\partial S}$ 
as,
\begin{equation}
 T= \frac{(S-2 \pi q^2)\sqrt{S+\pi q^2}}{4 \sqrt{\pi} S^2}.
\end{equation}
Fig(\ref{tempbardeen}) shows a smooth variation of temperature with respect to entropy, which excludes the chance of first order phase transition.  
We will now calculate the heat capacity, $C= T \frac{\partial S}{\partial T}$, of the black hole 
and is given by
\begin{equation}
 C= \frac{-2 S (2 \pi q^2-S)(\pi q^2 +S)}{8 \pi^2 q^4 + 4 \pi q^2 S - S^2}.
\end{equation}
\begin{figure}[h]
\centering
\includegraphics[scale=0.65]{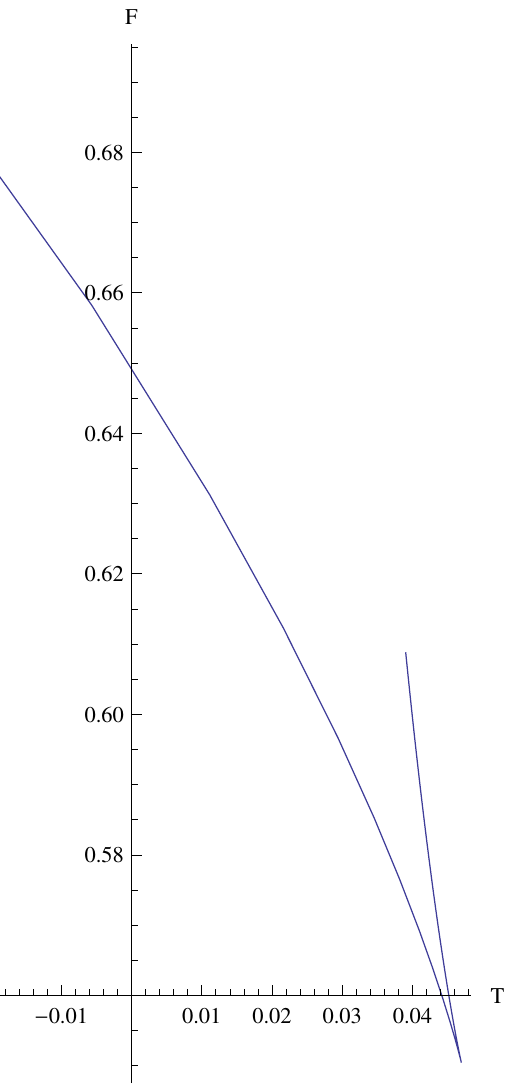}
\caption{Variation of free energy  with respect to temperature of Bardeen black hole.}
\label{freebardeen} 
\end{figure}
Then the Gibb's free energy, $F= M-TS$, is given by 
\begin{equation}
 F= \frac{( S+\pi q^2)}{2 \sqrt{\pi}S} -\frac{(S-2 \pi q^2)\sqrt{S+\pi q^2}}{4 \sqrt{\pi} S}.
\end{equation}
Now we study whether this black hole will undergo a second order phase transition or not. This can be done in two ways. 
The first method is by studying the variation of the heat capacity with entropy and we can see a discontinuity in heat capacity(Fig\ref{heatbardeen})
for a particular value of entropy( $S=4.1, q=0.5$). And we also note that heat capacity possesses a positive phase below this value of $S$ 
and a negative phase above this value of $S$.   
The second method is by studying the variation of free energy(F) with temperature(T). 

A second order phase transition is obvious for two reasons. First, of course the heat capacity shows an infinite discontinuity  (at $S=4.1$, where $q=0.5$)
and possesses both positive and negative phases. The positive phase exists for small values of $S$ and 
the black hole is stable only in this
region. The same result can also be seen from the parametric plot between the free energy and the temperature (Fig\ref{freebardeen}) which
shows a cusp type double point (at $T=0.046$, where $S=4.1$ and $q=0.5$). The $F$-$T$ variation  shows that
there are two branches of the curve, for one of the branches, the free energy decreases
with the increase of Hawking temperature to the minimum limit, while for the other branch
F increases rapidly with T. This behavior also signals a second order phase transition. (The numerical values are obtained from the corresponding graphs.)

 \subsection*{Ay\'{o}n-Beato and Garc\'{i}a black hole}

Ay\'{o}n-Beato and Garc\'{i}a (ABG) proposed a model of non-linear electrodynamics coupled to Einstein's gravity
to obtain a regular black hole as an exact solution\cite{abg1}. They proved that the Bardeen black hole was an exact solution
in a model of space-time with nonlinear electrodynamics.
It is also remarkable that all the nonlinear electrodynamics
satisfy  the zeroth and first laws of black hole mechanics\cite{rasheed}. 
Regular black hole solutions can be identified as exact
solutions to a model of nonlinear electrodynamics coupled to
Einstein gravity \cite{abg2,abg3,abg4,abg5}. 
The ABG solution is interpreted as a magnetically charged solution(it can be interpreted as the solution for a nonlinear magnetic monopole with 
mass M and charge q), that the corresponding 
potential V is taken in terms of the variation of the electric charge.
The metric of ABG black hole is,
\begin{equation}
ds^{2}=f(r)dt^{2}-\frac{dr^{2}}{f(r)}-r^{2}(d\theta^{2}+\sin^{2}\theta d\varphi^{2}),
\end{equation}
where
\begin{equation}
 f(r)= 1-\frac{2Mr^{2}}{(r^{2}+q^{2})^{\frac{3}{2}}}+ \frac{q^2 r^4}{(r^2 +q^2)^2}.
\end{equation}
By plotting $f(r)$ with respect to $r$, we can find the zeros of $f(r)$ and there by having information about event horizons.

\begin{figure}[h]
\centering
\includegraphics[scale=0.65]{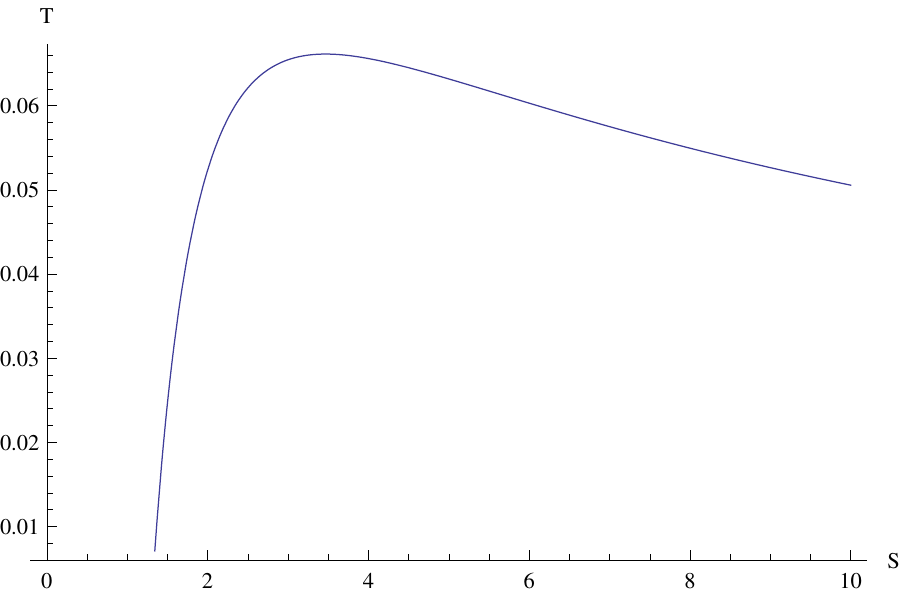}
\caption{Variation of temperature   with respect to entropy of ABG black hole.}
\label{ta}
\end{figure}

\begin{figure}[h]
\centering
\includegraphics[scale=0.65]{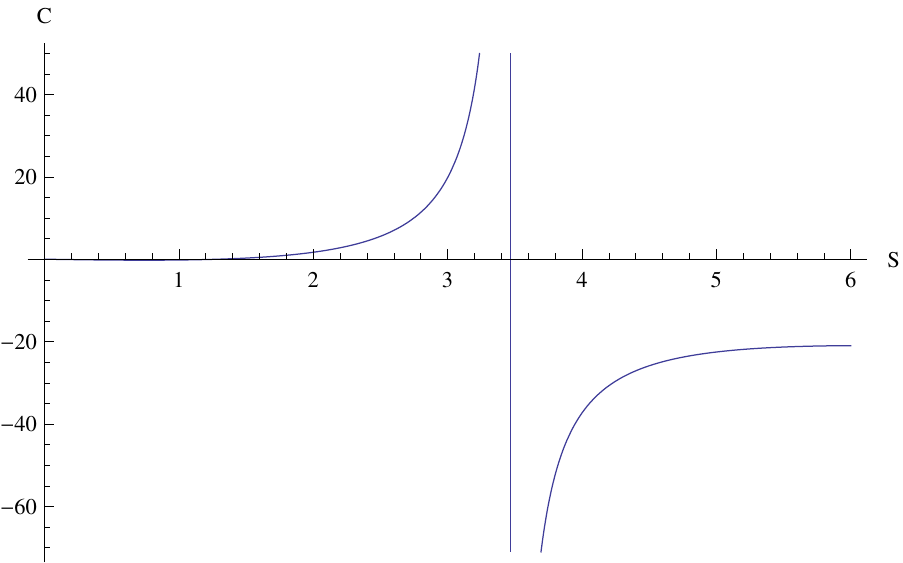}
\caption{Variation of heat capacity  with respect to entropy of ABG black hole.}
\label{ca}
\end{figure}

\begin{figure}[h]
\centering
\includegraphics[scale=0.65]{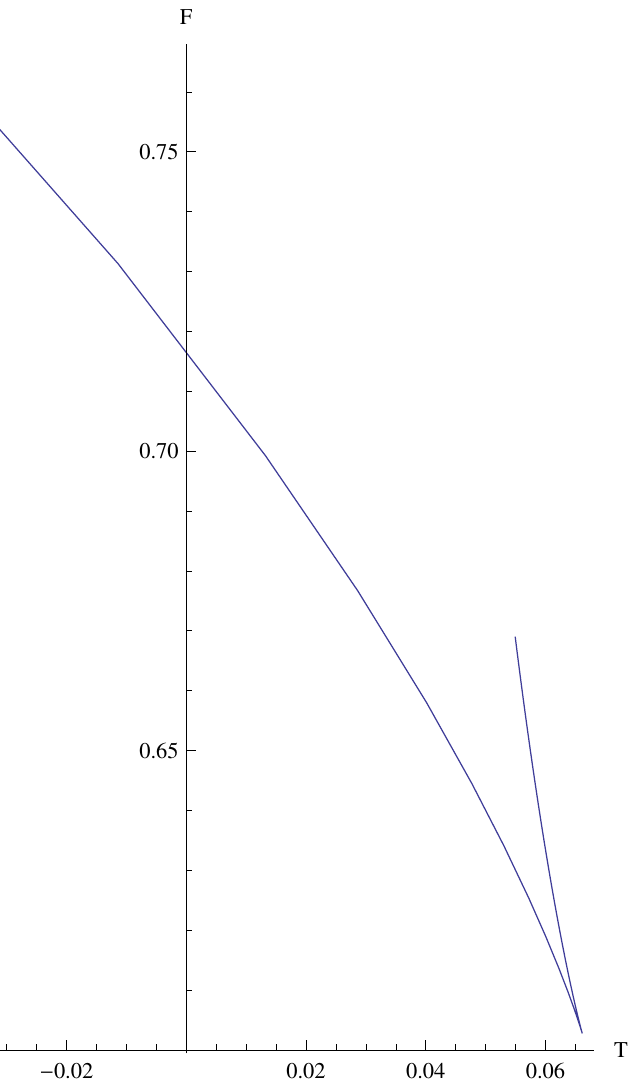}
\caption{Variation of free energy  with respect to temperature of ABG black hole.}
\label{fa}
\end{figure}

Proceeding as above, we can write the mass function as
\begin{equation}
 M = \frac{(q^4 + 2 q^2 r^2 + r^4 + q^2 r^4)}{(2 r^2 \sqrt{q^2 + r^2})}.
\end{equation}
Temperature can be written as
\begin{equation}
 T = \frac{(-2 \pi^3 q^6 - 3 \pi^2 q^4 S + 
  2\pi q^4 S^2 + (1 + q^2) S^3)}{
 4 \sqrt{\pi} S^2 (\pi q^2 + S)^{\frac{3}{2}}},
\end{equation}
and finally the heat capacity will be
\begin{equation}
C = \frac{2 S (\pi q^2 + S) (-2 \pi^3 q^6 - 3 \pi^2 q^4 S + 
    2 \pi q^4 S^2 + (1 + q^2) S^3)}{
 8 \pi^4 q^8 + 20 \pi^3 q^6 S + 15 \pi^2 q^4 S^2 + 
  2 \pi q^2 (1 - 2 q^2) S^3 - (1 + q^2) S^4}.
\end{equation}
The free energy also can be found out and is given by,
\begin{equation}
 F = \frac{4 \pi^3 q^6 + 9 \pi^2 q^4 s + 
  6 \pi q^2 s^2 + (1 + q^2) s^3}{
 4 \sqrt{\pi} s (\pi q^2 + s)^{\frac{3}{2}}}.
\end{equation}

In the case of Ay\'{o}n-Beato and Garc\'{i}a black hole, we have derived all the thermodynamic parameters, especially
temperature, mass, free energy and heat capacity. Here also we find that the temperature-entropy plot (Fig(\ref{ta})),
eliminates the chance of a first order phase transition,
whereas the heat capacity-entropy plot (Fig(\ref{ca})) shows a phase transition (at $S=3.46$, where $q=0.5$). 
And the similar cusp obtained in the free energy-temperature parametric plot
(at $T=0.066$, where $S=3.46$ and $q=0.5$ ), 
(Fig(\ref{fa})) confirms that ABG black hole also exhibits a second order phase transition
as shown by the regular black holes considered in this study.

 \subsection*{Hayward black hole.}

Hayward\cite{Hayward} obtained another regular black hole solution and  this space-time is helpful for us for further understanding of the 
behaviour of regular black holes.
The $f(r)$ is given in a minimal model, which satisfies the conditions, regarding the
validity of black holes and the existence of killing horizons. The cosmological constant $\Lambda$, is connected with $l$, as $\Lambda=\frac{3}{l^2}$.
The metric of Hayward black hole is given by,
\begin{equation}
ds^{2}=f(r)dt^{2}-\frac{dr^{2}}{f(r)}-r^{2}(d\theta^{2}+\sin^{2}\theta d\varphi^{2}),
\end{equation}
where
\begin{equation}
 f(r)=1-\frac{2Mr^2}{(r^3+2 \beta^2)},
\end{equation}
Here  $\beta^2=Ml^2$. 
The metric function of this black hole at large $r$ behaves as,
$$\lim_{r \rightarrow \infty} f(r) \rightarrow 1-\frac{2M}{r}+\mathcal{O}(\frac{1}{r^4}),$$
and at small $r,$ 
$$\lim_{r \rightarrow 0} f(r) \rightarrow 1-\frac{r^2}{l^2}+\mathcal{O}(r^5).$$ Thus this particular black hole solution
has well-defined asymptotic limits, namely it is Schwarzschild for $r \rightarrow \infty$ and de Sitter for $r \rightarrow0.$

 Mass of this Hayward black hole is given by(taking area law into account),
\begin{equation}
 M=\frac{S^{3/2}}{2\sqrt{\pi } \left(S-\pi l^2\right)}.
\end{equation}
Now the temperature is given by,
\begin{figure}[h]
\centering
\includegraphics[scale=0.65]{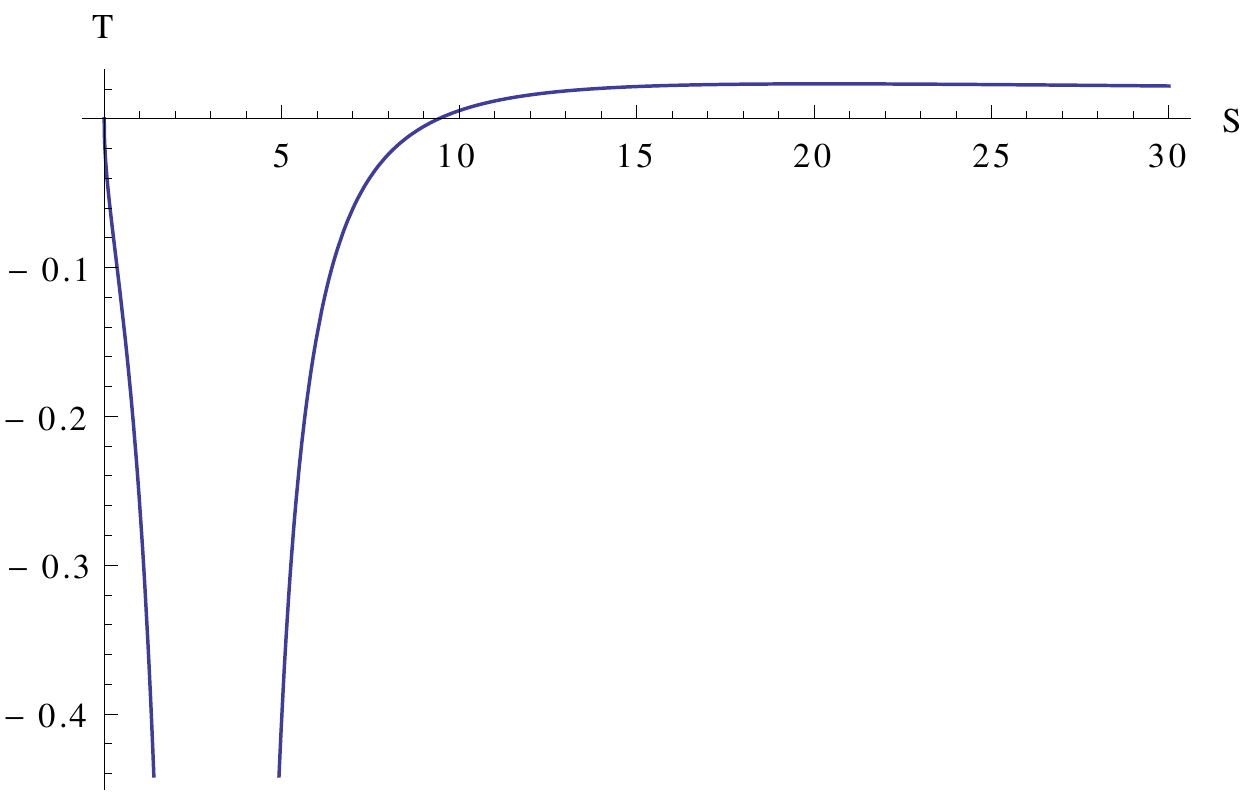}
\caption{ Variation of temperature of Hayward black hole  with respect to entropy.}
\label{th}
\end{figure}

\begin{equation}
 T=\frac{\sqrt{S} \left(S-3 \pi  l^2\right)}{4 \sqrt{\pi } \left(S-\pi
  l^2\right)^2}.
\end{equation}.
\begin{figure}[h]
\centering
\includegraphics[scale=0.65]{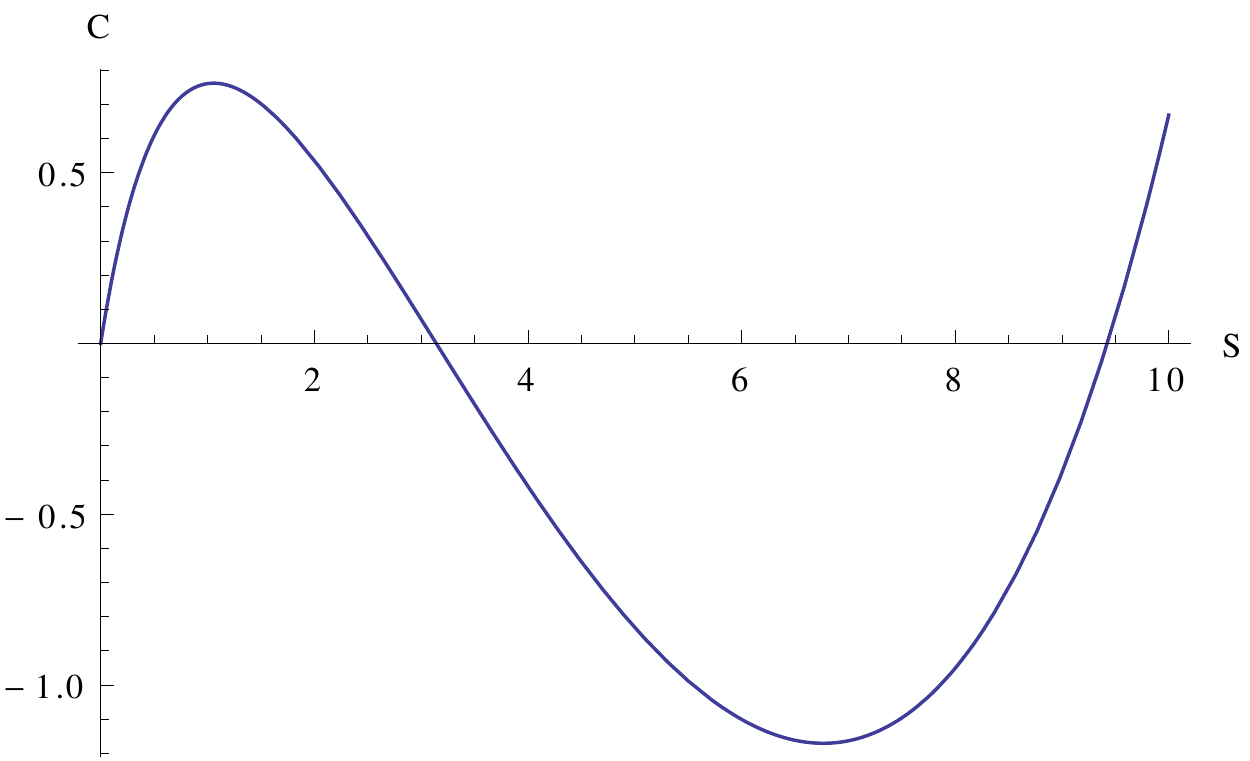}
\caption{ Variation of heat capacity of Hayward black hole  with respect to entropy(in the lower range of S).}
\label{ch1}
\end{figure}
The important observation here is that the temperature becomes negative below a certain value of S and
diverges. From the expression of temperature
it is obvious that it becomes singular at $S=\pi l^2$. The numerator of expression for temperature contains a factor $(S -3\pi l^2)$, which implies that 
$T$ will be negative for $S<3\pi l^2 $. So at a first look
it appears as a first order phase transition, but unfortunately it lies in the region which lacks any physical 
importance(negative temperature). Temperature is zero at $S= 3 \pi l^2$ and positive for $S>3\pi l^2 $. So this suggests a mass limit for Hayward black hole
formation, and hence a  minimum mass $M'$, corresponding to minimum value of  $S= 3 \pi l^2$. 
Thus the minimum value of mass required for the formation of Hayward black hole is given by
$$M'= \frac{3^{\frac{3}{2}}l}{4}$$. 
Now we look at the heat capacity of Hayward black hole, 
\begin{equation}
 C=\frac{2 S \left(\pi  l^2-S\right) \left(3 \pi  l^2-S\right)}{3 \pi
  ^2 l^4+6 \pi  l^2 S-S^2}.
\end{equation}
.
\begin{figure}[h]
\centering 
\includegraphics[scale=0.65]{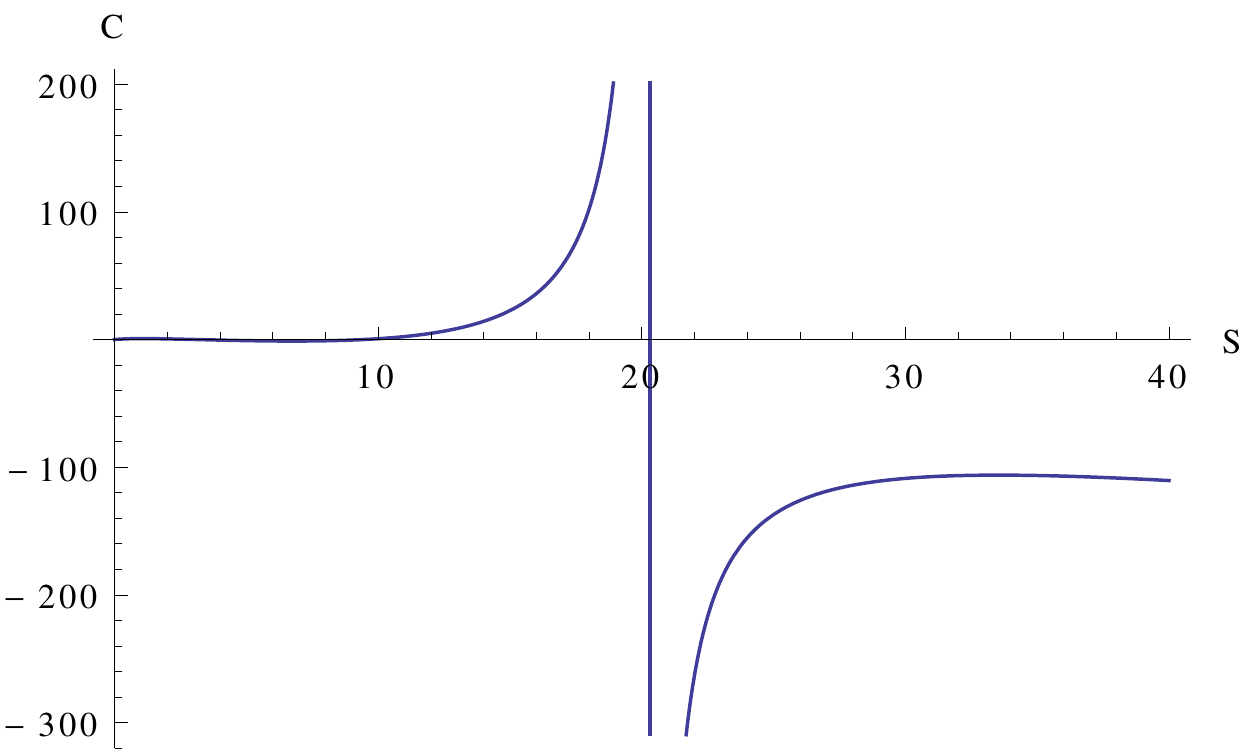}
\caption{ Variation of heat capacity of Hayward black hole  with respect to entropy(in the higher range of S).}
\label{ch2}
\end{figure}
From this figure we can see that the temperature diverges is at the value of  $S=3.14$.  
\begin{figure}[h]
\centering
\includegraphics[scale=0.65]{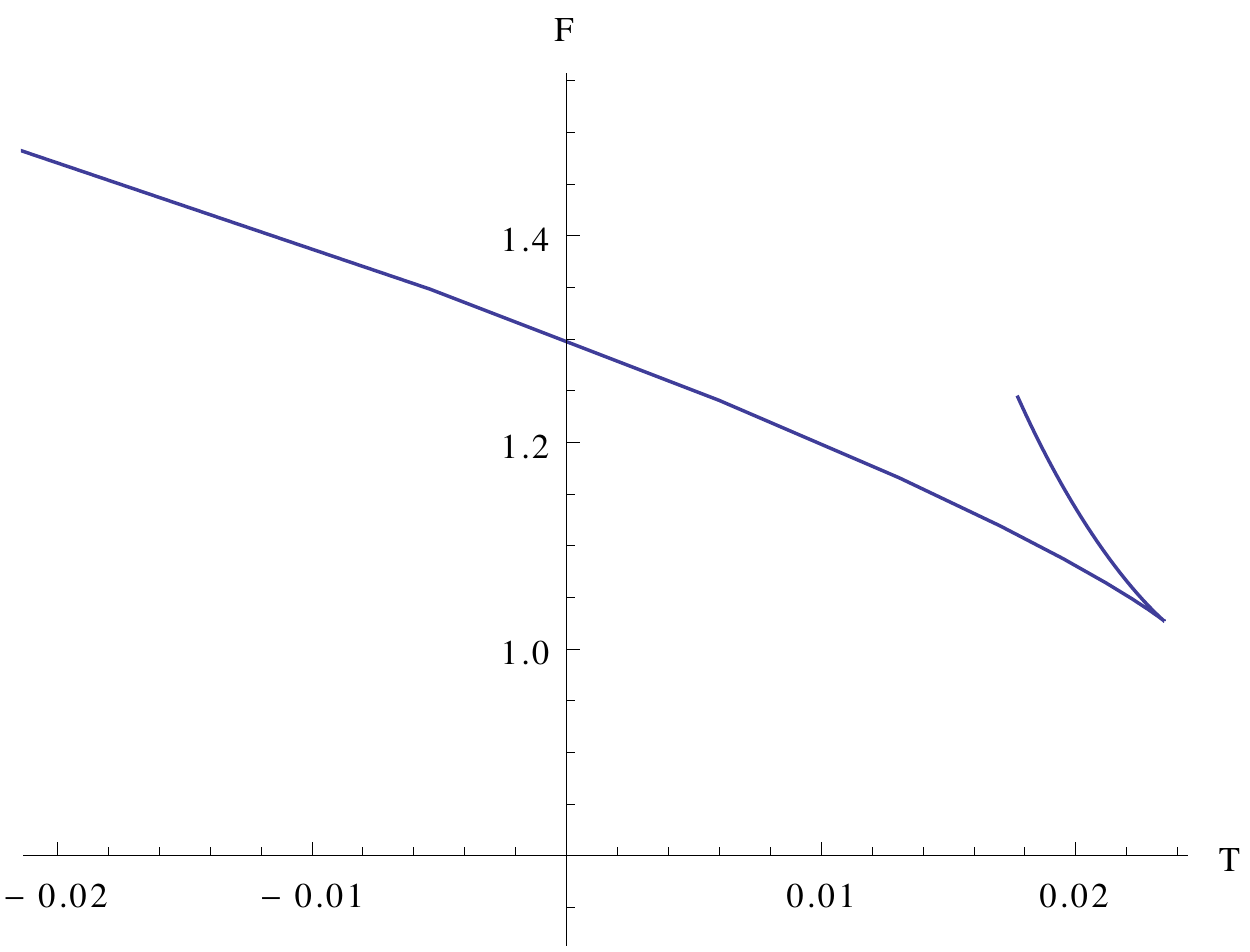}
\caption{Variation of free energy of Hayward black hole  with respect to temperature.}
\label{fh}
\end{figure}

On the other hand, the heat capacity undergoes the usual second order phase transition for large values of $S$. 
In the negative temperature region, the heat capacity shows a sinusoidal like variation. It is quite 
interesting to see that there 
is no singularity for heat capacity where the temperature has one. Heat capacity shows two kinds of behaviors. For lower values of S
(where the temperature shows a FOTPT),
the behavior is plotted in Fig(\ref{ch1}), and in the large $S$ values it shows SOTPT, as shown in Fig(\ref{ch2}), and the second order
phase transition occurs at a value of $S=20.28$ and the 
corresponding temperature is $T=0.0234$. Now 
we further confirm the second order phase transition with the help of free energy, 
\begin{equation}
 F=\frac{S^{3/2} (S+l^2 \pi )}{4 \sqrt{\pi } (S-l^2 \pi )^2}.
\end{equation}
The free energy-temperature plot is given in fig.8 and the cusp like double point occur for the same 
value of temperature $T=0.0234$, where the heat capacity diverges.

 \subsection*{Berej-Matyjasek-Trynieki-Wornowicz black hole}
  
BMTW black hole is yet another kind of regular black hole solution \cite{bereg2}, where the first order correction of the perturbative solution of the 
coupled equations of the quadratic gravity and nonlinear electrodynamics is constructed, with the zeroth order corrections coinciding with the 
ones given by the Ayon-Beato and Garzia black hole. The solution is parametrized by two integration constants and depends on two free parameters. 
Using the boundary conditions the integration constants are connected to the charge and the total mass of the system, 
whereas the free parameters are tuned to make the resultant line element regular at the center\cite{bereg2}.
The metric of BMTW black hole is,
\begin{equation}
ds^{2}=f(r)dt^{2}-\frac{dr^{2}}{f(r)}-r^{2}(d\theta^{2}+\sin^{2}\theta d\varphi^{2}),
\end{equation}
where
\begin{equation}
      f(r)=1-\frac{2 M \left(1-\tanh \left(\frac{\beta ^2}{2 M
  r}\right)\right)}{r}.
     \end{equation}
  Here also $\beta= Ml^2$, and $l$ is the cosmological constant related term.

 Mass of BMTW black hole can be written as a function of $S$(using the area law) and $l$ as,
 \begin{equation}
 M=\frac{\sqrt{S}}{\sqrt{\pi } \left(2-2 \tanh \left(\frac{\sqrt{\pi }
  l^2}{2 \sqrt{S}}\right)\right)},
 \end{equation}
   \begin{figure}[h]
\centering
\includegraphics[scale=0.65]{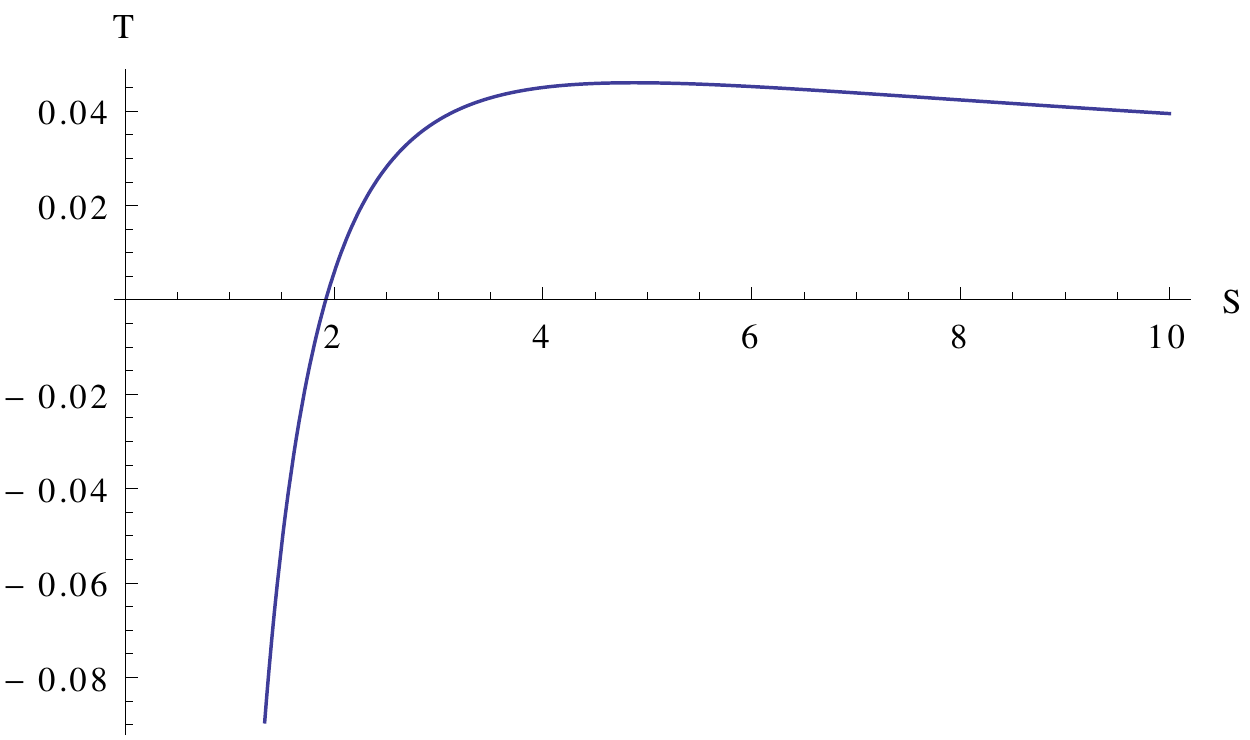}
\caption{Variation of temperature of BMTW black hole with respect to entropy.}
\label{tbg}
\end{figure}

  \begin{figure}[h]
\centering
\includegraphics[scale=0.65]{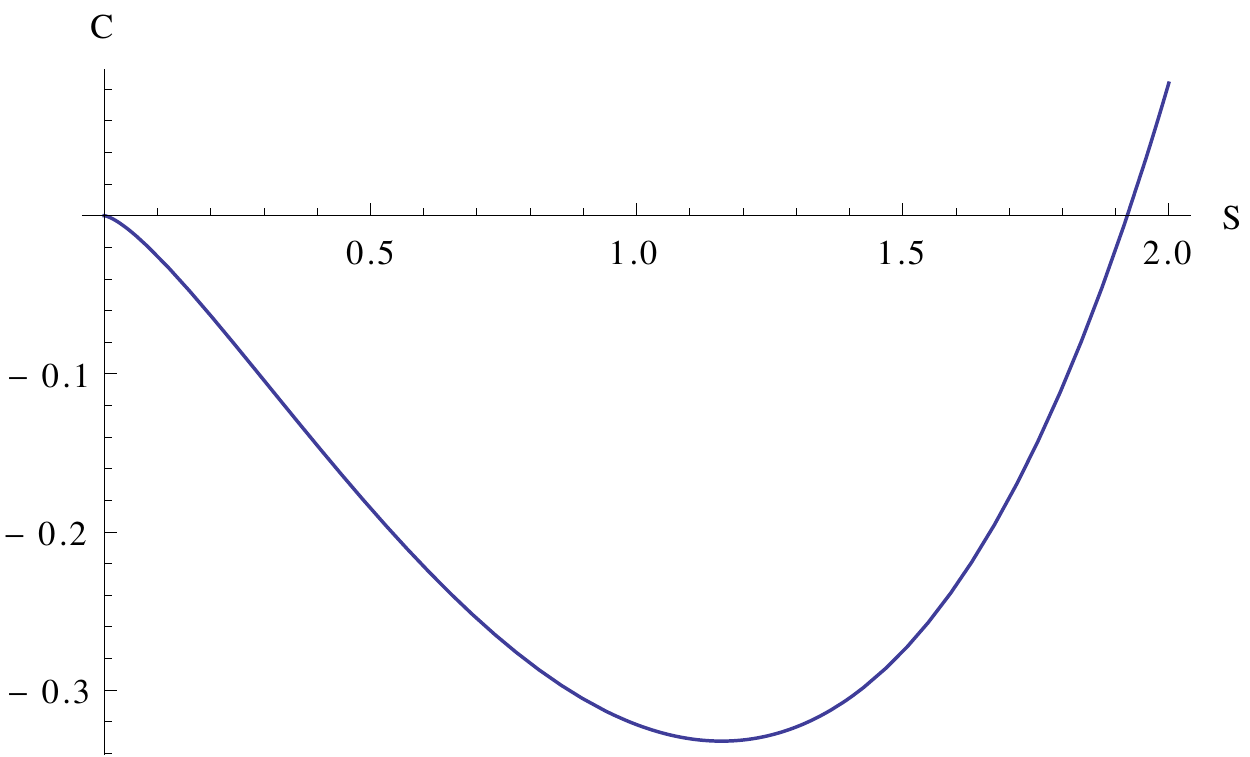}
\caption{Variation of heat capacity with respect to entropy for lower range of S of BMTW black hole.}
\label{cbg1}
\end{figure}

  \begin{figure}[h]
\centering
\includegraphics[scale=0.65]{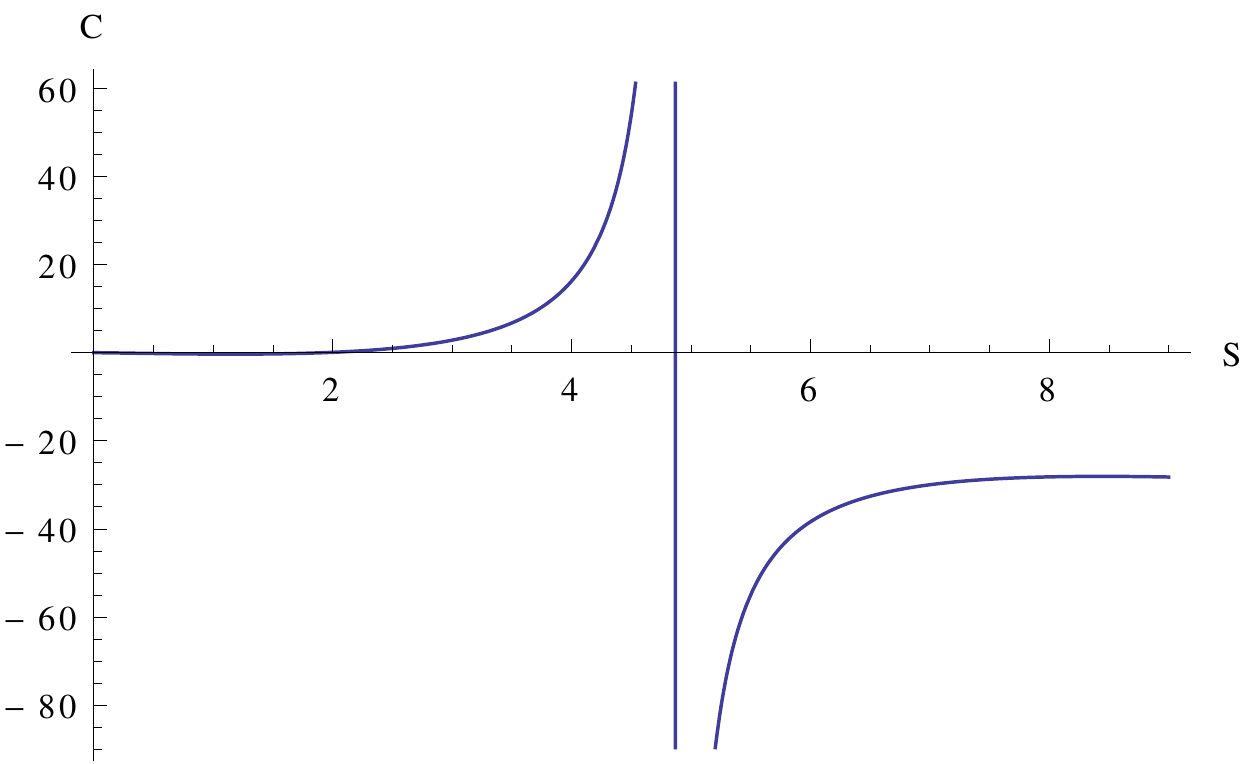}
\caption{Variation of heat capacity with respect to entropy for higher range of S of BMTW black hole.}
\label{cbg2}
\end{figure}
  and the temperature of BMTW black hole
\begin{equation}
 T=\frac{\sqrt{\pi } \left(-l^2\right)
  \text{sech}^2\left(\frac{\sqrt{\pi } l^2}{2 \sqrt{S}}\right)-2
  \sqrt{S} \left(\tanh \left(\frac{\sqrt{\pi } l^2}{2
  \sqrt{S}}\right)-1\right)}{8 \sqrt{\pi } S \left(\tanh
  \left(\frac{\sqrt{\pi } l^2}{2 \sqrt{S}}\right)-1\right)^2}.
\end{equation}
Now we will look at the heat capacity of BMTW black hole, 
\begin{equation}
 C=-\frac{2 S^{3/2} \left(\sqrt{\pi } l^2 \tanh \left(\frac{\sqrt{\pi }
  l^2}{2 \sqrt{S}}\right)+\sqrt{\pi } l^2-2 \sqrt{S}\right)}{\pi
  l^4+\sqrt{\pi } l^2 \sqrt{S}+\left(\pi  l^4+\sqrt{\pi } l^2
  \sqrt{S}\right) \tanh \left(\frac{\sqrt{\pi } l^2}{2
  \sqrt{S}}\right)-2 S},
\end{equation}
and the free energy is given by
\begin{equation}
 F=\frac{\sqrt{\pi } l^2 \text{sech}^2\left(\frac{\sqrt{\pi } l^2}{2
  \sqrt{S}}\right)-2 \sqrt{S} \left(\tanh \left(\frac{\sqrt{\pi }
  l^2}{2 \sqrt{S}}\right)-1\right)}{8 \sqrt{\pi } \left(\tanh
  \left(\frac{\sqrt{\pi } l^2}{2 \sqrt{S}}\right)-1\right)^2}.
\end{equation}
  \begin{figure}[h]
\centering
\includegraphics[scale=0.65]{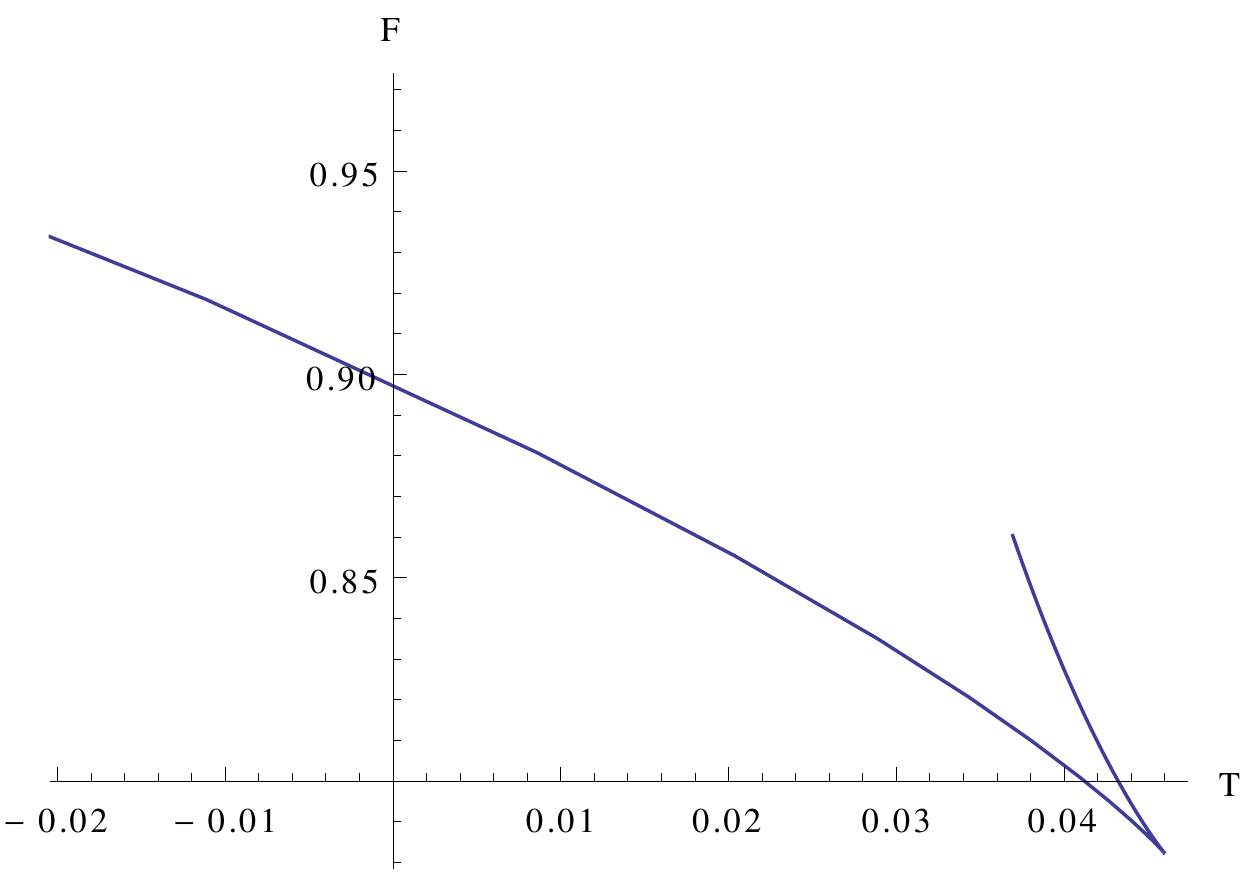}
\caption{Variation of free energy with respect to temperature of BMTW black hole.}
\label{fbg}
\end{figure}

Variation of temperature with entropy of  BMTW black hole is plotted in Fig(\ref{tbg}). 
There is no discontinuity in the temperature-entropy plot and thus we discard the chance of FOTPT. And a critical size of BMTW black hole
can be obtained from the $T=0$ point at which 
the value of entropy is $S=1.928$. This is the minimum possible entropy(in turn a minimum possible horizon radius) for a BMTW black hole to form.
Thus the unphysical regime of temperature, that is for the values $S<1.928$, is being reflected in the heat capacity(Fig(\ref{cbg1})),
for the lower values of $S$, heat capacity also goes through zero, from negative to positive, 
without a discontinuity. For large values of $S$, we could easily identify the SOTPT. 
The heat capacity here showing a SOTPT, at a particular value of 
$S=4.855$(Fig(\ref{cbg2})), and the same value of  $T=0.046$ can also be obtained from the cusp like double
point in the free energy- temperature plot (Fig(\ref{fbg})).

\section{Geometrothermodynamics}

Now we will investigate the thermodynamic properties of regular black holes using the concepts of Geometrothermodynamics (GTD) \cite{q1,q2,q3}
In order to describe a thermodynamic system with $n$ degrees of freedom, 
the first step is to think of a thermodynamic phase space which
is defined mathematically as a Riemannian contact
manifold$(\mathcal{T} , \Theta, G)$, where $\mathcal{T}$ is a $2n + 1$ dimensional manifold, 
$\Theta$ defines a contact structure on $\mathcal{T}$ and $G$ is a Legendre invariant metric on $\mathcal{T}$.
The pair $( \mathcal{T}, \Theta)$ is called a contact manifold\cite{hermann} only if $\mathcal{T}$ is
differentiable and $\Theta$ satisfies the condition $\Theta \wedge (d \Theta)^{n} \neq 0$, 
which actually preserves the essential Legendre invariance while making the conformal transformations.
The space of equilibrium states is an $n$ dimensional manifold $(\mathcal{E}, g)$, where $\mathcal{E} \subset \mathcal{T}$ is defined by a 
smooth mapping
$\phi : \mathcal{E} \rightarrow \mathcal{T}$ such that the pullback $\phi^* (\Theta) = 0$, and a
Riemannian structure $g$ is induced naturally in the $\mathcal{E}$ by
means of $g = \phi^{(G)} $. It is then expected in GTD that
the physical properties of a thermodynamic system in
a state of equilibrium can be described in terms of the
geometric properties of the corresponding space of equilibrium states $\mathcal{E}$. The smooth mapping can be read in
terms of coordinates as, $\phi:(E^{a}) \rightarrow (\Phi, E^{a} , I^{a} )$ with
$\Phi$ representing the thermodynamic potential, $E^{a}$ and
$I^{a}$ representing the extensive and intensive thermodynamic
variables respectively. If the condition $\phi^* (\Theta) = 0$ is satisfied, we can find,
\begin{equation}
  d\Phi = \delta_{ab}I^{a}dE^{b}\ , \quad \frac{\partial \Phi}{\partial E^{a}}=\delta_{ab}I^{b}\ .
\label{firstlaw}
\end{equation}
The first one of these equations corresponds to the first law of thermodynamics, whereas the second one is usually 
known as the condition for thermodynamic equilibrium\cite{callen}.

Legendre invariance guarantees that the geometric
properties of $G$ do not depend on the thermodynamic
potential used in its construction. Quevedo\cite{q1} introduced the idea and constructed a general form for
the Legendre invariant metric. The general choice of GTD metric is as follows
\begin{equation}
 g=\phi^*(G)= \left(E^{c}\frac{\partial{\Phi}}{\partial{E^{c}}}\right)
\left(\eta_{ab}\delta^{bc}\frac{\partial^{2}\Phi}{\partial {E^{c}}\partial{E^{d}}} dE^a dE^d \right).
\label{gtdmetric}
\end{equation}

The thermodynamic geometry of a black hole is still a most fascinating subject and there are many unresolved issues in black hole thermodynamics. Using this method 
we could solve for phase transition shown by the system. The earlier studies show that the thermodynamic stability 
of systems depend on the potential we have chosen.
This contradiction has been removed by using new Legendre invariant metric introduced in the GTD. In this
paper we describe the phase transition in terms of curvature singularities.

\subsection*{Bardeen black hole}
\begin{figure}[h]
\centering
\includegraphics[scale=0.65]{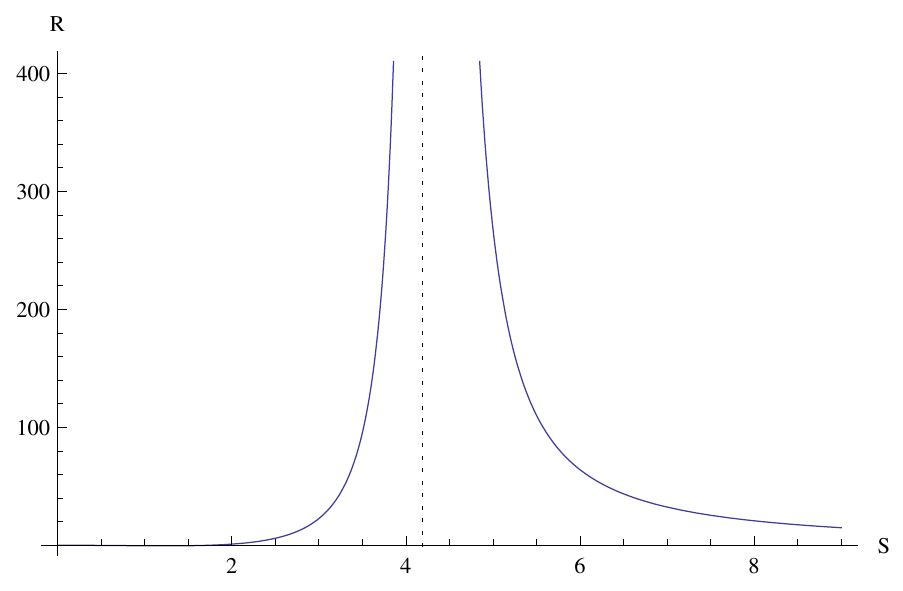}
\caption{\label{fig:orlfo} Variation of curvature scalar of Bardeen black hole with respect to entropy.}
\label{curvebardeen}
\end{figure}

We have $M$ as $M(S,q)$. Thus using (\ref{gtdmetric}) we could write the GTD metric as
\begin{equation}
         g^{GTD}=(S M_S + q M_q )
            \left[ {\begin{array}{cc}
             -M_{SS} & 0 \\
             0 & M_{qq} \\
             \end{array} } \right].
        \end{equation}
Now 
$$M_{SS}=\frac{8 \pi^2 q^4 + 4 \pi q^2 S- S^2}{8 S^3 \sqrt{\pi^2 q^2 + \pi S}}$$
and
$$M_{qq}= \frac{3\sqrt{\pi}(2 \pi q^2 +S)}{2 S \sqrt{\pi q^2 +S}}$$
The curvature scalar gets the form
\begin{equation}
 R^{GTD}= \frac{f(S,q)}{(2 q^2 \pi + S)^2 (4 q^2 \pi + S)^3 (-8 q^4 \pi^2 - 
   4 q^2 \pi S + S^2)^2},
\end{equation}
where $f(S,q)$ is a complicated expression with less
physical interest.
We have found the curvature and plotted it (Fig\ref{curvebardeen}). We can see that $R$ shows a singular behavior for the values of    
$S=4.1$ and $q=0.5$. From our earlier study we can see that, the values of $S$ and $q$ coincide with the values of $S$ and $q$ for which the 
heat capacity becomes divergent.

Now we can say that a singularity in specific heat corresponds to a singularity in scalar 
curvature for the same value of entropy.

 \subsection*{ABG black hole}
\begin{figure}[h]
\centering
\includegraphics[scale=0.65]{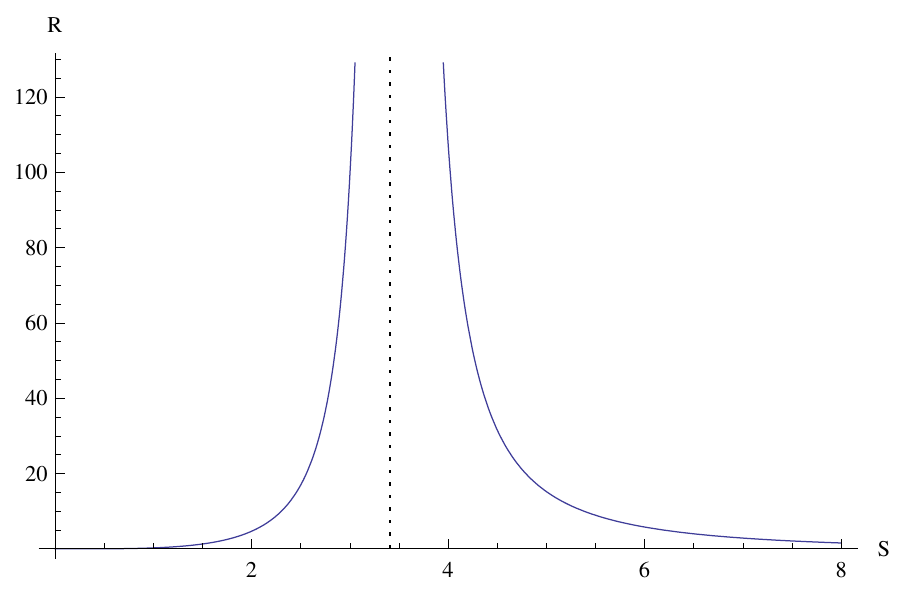}
\caption{Variation of curvature scalar of ABG black hole with respect to entropy.}
\label{ra}
\end{figure}
Now we analyze the state space geometry of this black hole. We could find the GTD metric in the same manner
as is done for other regular black holes. 
We have $M$ as $M(S,q)$, thus we could write the GTD metric as
\begin{equation}
         g^{GTD}=(S M_S + q M_q )
            \left[ {\begin{array}{cc}
             -M_{SS} & 0 \\
             0 & M_{qq} \\
             \end{array} } \right].
        \end{equation}
where
        $$M_{SS}=\frac{8 \pi^4 q^8 + 20 \pi^3 q^6 s + 15 \pi^2 q^4 s^2 + 
 2 \pi q^2 (1 - 2 q^2) s^3 - (1 + 
    q^2) s^4}{(8 \sqrt{\pi} s^3 (\pi q^2 + s)^\frac{5}{2}}$$,
and 
$$M_{qq}=\frac{6 \pi^4 q^6 + 15 \pi^3 q^4 s + 
 12 \pi^2 q^2 s^2 - \pi (-3 + q^2) s^3 + 
 2 s^4}{2 \sqrt{\pi} s (\pi q^2 + s)^\frac{5}{2}}$$.
And the curvature scalar of GTD metric of ABG black hole is,
we obtain
\begin{equation}
 R^{GTD}= \frac{f^{GTD}}{((8 q^8 \pi^4 + 20 q^6 \pi^3 s + q^4 \pi (15 \pi - 4 s) s^2 + 
    q^2 (2 \pi - s) s^3 - s^4)^2 (6 q^6 \pi^4 + 15 q^4 \pi^3 s +
     q^2 \pi (12 \pi - s) s^2 + 
    s^3 (3 \pi + 2 s))^2 (4 q^6 \pi^3 + s^3 + 
    q^4 \pi s (9 \pi + 4 s) + q^2 s^2 (6 \pi + 5 s))^3)}.
\end{equation}
We plot the curvature scalar-entropy graph (Fig(\ref{ra})) and find that the singularity matches with that of heat capacity (at $S=3.46$, where $q=0.5$).

 \subsection*{Hayward black hole}
No we will look at the GTD of Hayward black hole. The GTD metric can be derived from $M(S,l)$;

\begin{equation}
         g^{GTD}=(S M_S + l M_l )
            \left[ {\begin{array}{cc}
             -M_{SS} & 0 \\
             0 & M_{ll} \\
             \end{array} } \right].
        \end{equation}
        
        Where $$M_{SS}=\frac{-3 \pi ^2 l^4-6 \pi  l^2 S+S^2}{8 \sqrt{\pi } \sqrt{S}
  \left(\pi  l^2-S\right)^3}$$  and
        
 $$M_{ll}=\frac{\sqrt{\pi } S^{3/2} \left(3 \pi  l^2+S\right)}{\left(S-\pi
  l^2\right)^3}$$. 
  
  \begin{figure}[h]
\centering
\includegraphics[scale=0.65]{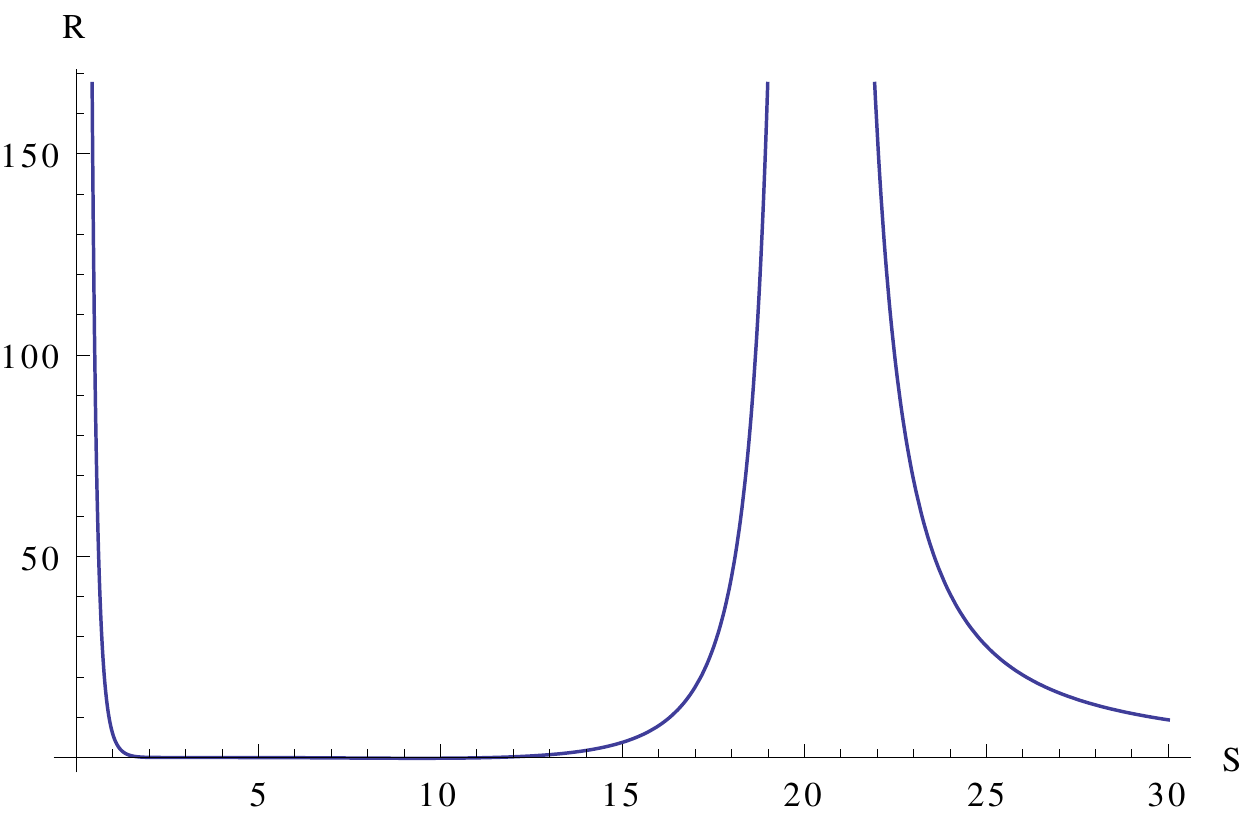}
\caption{curvature scalar of Hayward black hole  showing singularities.}
\label{rh}
\end{figure}
  
  And finally the curvature scalar reads like,
  \begin{equation}
   R^{GTD}=\frac{f^{GTD}}{s^2 \left(\pi  J^2+s\right)^3 \left(3 \pi  J^2+s\right)^2 \left(-3 \pi ^2 J^4-6 \pi  J^2 s+s^2\right)^2}.
  \end{equation}
 From Fig(\ref{rh})the curvature scalar becomes singular at $S=20.28$. We saw earlier that  the heat capacity showed an infinite discontinuity for
 the same  value of $S$.

 \subsection*{BMTW black hole}
 The GTD of BMTW black hole is given by
 \begin{equation}
           g^{GTD}=(S M_S + l M_l )
            \left[ {\begin{array}{cc}
             -M_{SS} & 0 \\
             0 & M_{ll} \\
             \end{array} } \right].
 \end{equation}
here also we can explicitly define $M_{SS}$ and  $M_{ll}$;
$$ M_{SS}=-\frac{\left(\pi  l^4+\sqrt{\pi } l^2 \sqrt{s}+\left(\pi
  l^4+\sqrt{\pi } l^2 \sqrt{s}\right) \tanh \left(\frac{\sqrt{\pi }
  l^2}{2 \sqrt{s}}\right)-2 s\right)
  \text{sech}^2\left(\frac{\sqrt{\pi } l^2}{2 \sqrt{s}}\right)
  \left(\cosh \left(\frac{\sqrt{\pi } l^2}{\sqrt{s}}\right)-\sinh
  \left(\frac{\sqrt{\pi } l^2}{\sqrt{s}}\right)\right)}{16
  \sqrt{\pi } s^{5/2} \left(\tanh \left(\frac{\sqrt{\pi } l^2}{2
  \sqrt{s}}\right)-1\right)^3}$$.
  And
  $$M_{ll}=\frac{\left(2 \sqrt{\pi } l^2+\sqrt{s}\right) \left(\sinh
  \left(\frac{\sqrt{\pi } l^2}{\sqrt{s}}\right)+\cosh
  \left(\frac{\sqrt{\pi } l^2}{\sqrt{s}}\right)\right)}{2 \sqrt{s}}$$
  
  and the curvature scalar is 
  \begin{equation}
   R^{GTD}=\frac{1}{\left(2 \sqrt{\pi } J^2+\sqrt{s}\right)^2 \left(3 \sqrt{\pi
  } J^2 \tanh \left(\frac{\sqrt{\pi } J^2}{2 \sqrt{s}}\right)+3
  \sqrt{\pi } J^2+2 \sqrt{s}\right)^3 \left(\pi  J^4+\sqrt{\pi }
  J^2 \sqrt{s}+\left(\pi  J^4+\sqrt{\pi } J^2 \sqrt{s}\right) \tanh
  \left(\frac{\sqrt{\pi } J^2}{2 \sqrt{s}}\right)-2 s\right)^2
  \left(\sinh \left(\frac{3 \sqrt{\pi } J^2}{2
  \sqrt{s}}\right)+\cosh \left(\frac{3 \sqrt{\pi } J^2}{2
  \sqrt{s}}\right)\right)}
  \end{equation}
  
    \begin{figure}[h]
\centering
\includegraphics[scale=0.65]{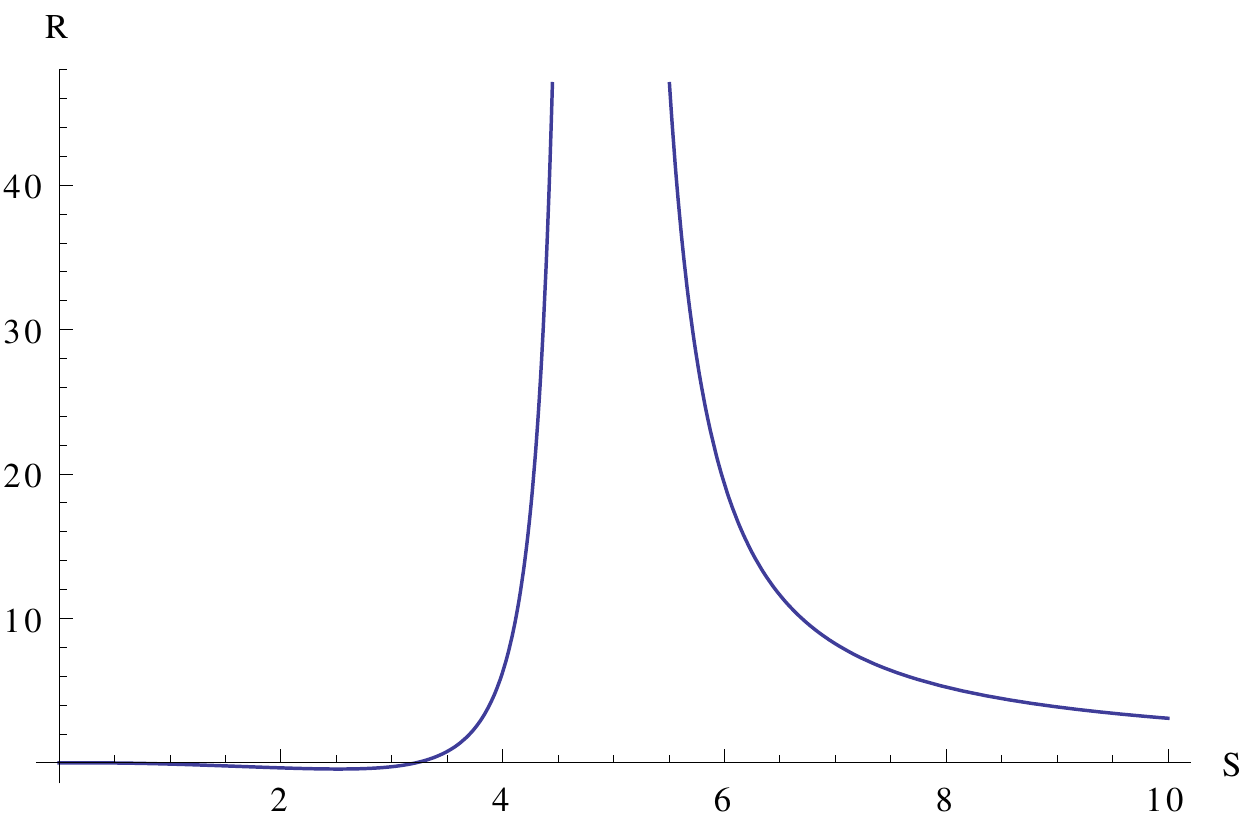}
\caption{Variation of curvature scalar of BMTW black hole with entropy}
\label{rbg}
\end{figure}
Here also the curvature scalar has got the singularity(Fig(\ref{rbg}) matching with that of SOTPT we observed in the case of BMTW black hole, at $S=1.928$.

\section{Results and Conclusion}

In this work we have studied the thermodynamics of regular black holes introduced by Bardeen, Hayward, Ay\'{o}n-Beato and Garc\'{i}a and BMTW.
We have derived the thermodynamic quantities, and plotted their variations with 
respect to entropy. From the temperature-entropy diagram we have eliminated the possibility of a first order phase transition in 
all cases. In the case of Hayward black hole the temperature-entropy plot shows 'FOTPT-like' behaviour, but unfortunately it is in the unphysical 
regime of temperature, where temperature 
is negative. We have established existence of second order phase transitions in all these regular black holes in two ways.
Both heat capacity-entropy plot and the parametric plot between 
free energy and temperature (a cusp like double point in $F-T$ graph), ensure
the existence of singularity. We further studied the Legendre invariant metric of these regular black holes. 
The particular thermodynamic
metrics delivered by GTD also describe the 
thermodynamic behavior of the corresponding black hole configurations. Scalar curvature related to the thermodynamic metric 
is non-zero and its singularities 
reproduce the phase transition structure as in thermodynamic studies. Thus the differential geometric formalism, whose objective is
to describe the thermodynamic behaviour of physical systems in  an invariant manner, could explain the thermodynamics of regular black holes.

\section*{Acknowledgements}
TR wishes to thank UGC, New Delhi for financial support
under RFSMS scheme. JS  and VCK Wishes to thank UGC, for the financial support through a major research project sanctioned to
VCK. VCK also wishes to acknowledge Associateship of
IUCAA, Pune, India.


\begin{thebibliography}{15}



\bibitem{Bardeen}J. Bardeen, Proceedings of GR5, Tiflis, U. S. S.
R


\bibitem{abg1}
E.Ay\'{o}n-Beato and A. Garc\'{i}a, Phys. Rev. Let., \textbf{80}, 5056 (1998)


\bibitem{Hayward}
Sean A. Hayward, Phys. Rev. Lett. \textbf{96}, 031103 (2006)



\bibitem{bereg2}
Waldemar Berej, Jerzy Matyjasek, Dariusz Tryniecki, Mariusz Woronowicz, Gen. Relativ. Gravit. \textbf{38}(5): 885–906 (2006) 



\bibitem{borde1}
A. Borde, Phys. Rev. D,  \textbf{50}, 3692 (1994)

\bibitem{borde2}
A. Borde, Phys. Rev. D,  \textbf{55}, 7615 (1997)



\bibitem{BCH} J. M. Bardeen, B. Carter and S. W. Hawking, Commun. Math. Phys. \textbf{31}, 161 (1973)


\bibitem{Z}
S. Zhou, J. Chen, Y. Wang, arxiv:1112.5909


\bibitem{E}
E. F. Eiroa, C. M. sendra, Clas. Quantum. Grav, \textbf{28}, 085008 (2011)


\bibitem{mor}
C. Moreno, O. Sarbach, Phys. Rev. D \textbf{67}, 024028 (2003)



\bibitem{fer}
S. Fernando, J.Correa, Phys. Rev. D \textbf{86}, 064039 (2012)


\bibitem{akbar}
M. Akbar, Nema Salem, S. A. Hussein, Chin. Phys. Lett. \textbf{29}, 7, 070401 (2012)




\bibitem{hermann}
R. Hermann., \textit{Geometry, physics and systems}., Marcel Dekker., New York (1973).


\bibitem{mrugala} 
R. Mrugala., Physica A (Amsterdam)., \textbf{125}, 631 (1984).



\bibitem{weinhold} 
F. Weinhold., J. Chem. Phys., \textbf{63}, 2479 (1975).

\bibitem{ruppeiner}
G. Ruppeiner., Phys. Rev. A., \textbf{20}, 1608 (1979).


\bibitem{q1}
H. Quevedo., J. Math. Phys (N.Y)., \textbf{48}, 013506 (2007).


\bibitem{q2} 
H. Quevedo., Gen. Relativ. Gravit., \textbf{40}, 971 (2008).

\bibitem{q3}
H. Quevedo and A. Vazquez, in \textit{Recent Developments in Gravitation and Cosmology}, edited by Alfredo Macias, Claus L\"{a}mmerzahl, and Abel 
Camacho, AIP Conf. Proc. No. 977 (AIP, New York, 2008).

\bibitem{tj1}
Jishnu Suresh, R.Tharanath,Nijo Varghese, V.C. Kuriakose, Eur. Phys. J. C \textbf{74}, 2819 (2014) 

\bibitem{tj2}
       R.Tharanath, Jishnu Suresh, Nijo Varghese, V.C. Kuriakose, Gen. Relativ. Gravit. DOI:10.1007/s10714-014-1743-x

\bibitem{callen}
Callen. H. B. \textit{Thermodynamics and an Introduction to Thermostatics.} Wiley, Newyork, (1985).



\bibitem{rasheed}
D. A. Rasheed, Non-Linear electrodynamics: Zeroth and first law of Black hole mechanics, hep-th/9901133, 1999.


\bibitem{abg2}
E.Ay\'{o}n-Beato and A. Garc\'{i}a, Phys. Let.B, \textbf{464}, 25 (1999)


\bibitem{abg3}
E.Ay\'{o}n-Beato and A. Garc\'{i}a, Gen. Relativ. Grav., \textbf{31}, 629 (1999)

\bibitem{abg4}
E.Ay\'{o}n-Beato and A. Garc\'{i}a, Phys. Let.B, \textbf{493}, 149 (2000)


\bibitem{abg5}
E.Ay\'{o}n-Beato and A. Garc\'{i}a, Gen. Relativ. Grav., \textbf{37}, 635 (2005)


\bibitem{tharanath}
R. Tharanath, V. C. Kuriakose, Mod. Phys. Lett. A \textbf{28}, 4 (2013).


\bibitem{tharanath2}
R. Tharanath, Nijo Varghese and V. C. Kuriakose, Mod. Phys. Lett. A \textbf{29}, 11(2014).

\end{thebibliography}
\end{document}